# A Novel Four-Stage Synchronized Chaotic Map: Design and Statistical Characterization

Ricardo Francisco Martinez-Gonzalez

Tecnológico Nacional de México - IT Veracruz, Veracruz, Veracruz, Mexico

## Abstract

*Digital implementations of chaotic systems often suffer from inherent degradation, limiting their long-term performance and statistical quality. To address this challenge, we propose a novel four-stage synchronized piecewise linear chaotic map. This new map is meticulously designed with four independent segments, each possessing its own control parameters, specifically engineered to mitigate the natural degradation observed in digitally realized dynamical systems. We characterize its behavior using established tools from nonlinear dynamics, including bifurcation diagrams and graphical analysis, which provide a comprehensive qualitative understanding of its complex dynamics.*

*To rigorously validate the statistical features of the generated sequences, we employed the National Institute of Standards and Technology (NIST) statistical testing suite. A substantial 100 MB dataset, comprising sequences produced by the proposed map, was generated via a Matlab script and subjected to this rigorous battery of tests. Our results demonstrate that the proposed map exhibits superior statistical properties compared to the classic Bernoulli map, successfully passing all NIST tests where the traditional map did not. This research confirms the proposed map's potential as a robust and high-quality source for chaotic sequence generation.*

## Keywords

*Chaotic maps, Synchronized stages, Statistical validation, Degradation mitigation*

## 1. Introduction

For several years, researchers have extensively studied the applications of chaos [1-4]. Chaos plays a fundamental role in various fields, including communications [5], cryptography [6], steganography [7], and other security services [8]. It is also a crucial component in mathematical [9] and economic simulations [10]. This diversity of applications stems from chaos's ability to mimic random noise [11], its hypersensitivity to initial conditions [12], its broad Fourier transform spectra, and the fractal properties of its motion in the phase plane [13].

To properly assess the quality of the "randomness" generated by chaotic systems, especially for cryptographic and security applications, it is essential to employ rigorous statistical tests. The National Institute of Standards and Technology (NIST) provides a comprehensive suite of statistical tests for randomness (NIST SP 800-22) [14]. These tests are widely accepted in the scientific community and are designed to detect various types of non-randomness in binary sequences, making them an indispensable tool for validating the suitability of chaotic generators for sensitive applications.

Numerous maps exist for generating chaos; however, they can be classified by the number of iterated variables or dimensions [15, 16]. Generally, maps with fewer dimensions are simpler to implement. Moreover, one-dimensional maps have already demonstrated excellent chaotic features [17], making them a good option for chaos generation [18].

However, when chaotic systems are implemented in digital platforms, they inevitably suffer from digital degradation due to finite precision effects and limited computational resources. This degradation can lead to a significant loss of chaotic properties, such as periodicity, shrinkage of the phase space, and a reduction in unpredictability, which can compromise the security and performance of chaotic applications [19, 20, 21]. Addressing these limitations is a critical challenge in the practical deployment of chaotic systems.

One-dimensional maps can be sub-classified based on the form of their characteristic curve. If it consists of segments, it is considered discontinuous; conversely, if it consists of a single segment, it is continuous [22]. Piecewise linear maps are considered discontinuous and have no theoretical limit on the number of segments [23], although authors typically restrict them to positive values. Our proposal aligns with the approach of switching multiple chaotic systems, as it comprises four similar but independent mapping functions. To lay the groundwork for our proposed system, we first review some of the well-established one-dimensional chaotic maps that inspire its design.

## 1.1. One-Dimensional Chaotic Maps

One-dimensional maps are foundational in chaos theory due to their relative simplicity and ability to exhibit complex dynamics, making them suitable for various applications despite their low dimensionality. Several classic one-dimensional maps serve as benchmarks and are widely studied:

- Bernoulli Map: Also known as the doubling map, the Bernoulli map is a fundamental example of a chaotic map. Defined as $x_{n+1}=(2x_n)(\text{mod}1)$ [24], it stretches and folds the unit interval [0,1], demonstrating properties like ergodicity and mixing, which are crucial for generating sequences with good randomness properties. Its simplicity allows for straightforward analysis of its chaotic behavior.
- Logistic Map: Perhaps the most famous and extensively studied one-dimensional map, the Logistic map is given by $x_{n+1}=rx_n(1-x_n)$ [17]. It famously demonstrates how simple non-linear equations can lead to highly complex, chaotic behavior through a series of period-doubling bifurcations as the parameter r is varied. For r=4, the map exhibits full chaos over the interval [0,1]. Its applications range from population dynamics to pseudorandom number generation.
- Tent Map: Characterized by its piecewise linear nature, the Tent map is defined as $x_{n+1}=\mu\min(x_n,1-x_n)$ [11]. For µ=2, it maps the interval [0,1] onto itself, producing chaotic sequences. Its linear segments make it easier to analyze some of its properties compared to non-linear maps like the Logistic map, while still possessing strong chaotic characteristics suitable for practical applications.
- Sine Map: Another non-linear one-dimensional chaotic map, the Sine map is defined by $x_{n+1}=(A/\pi)\sin(\pi x_n)$ [25]. For specific values of parameter A (e.g., A=4), it exhibits chaotic behavior similar to the Logistic map, mapping the interval [0,1] to itself. Its smooth, sinusoidal characteristic curve contributes to its rich dynamic properties.

These one-dimensional maps, along with many others, provide the theoretical basis and practical tools for generating chaotic sequences used in diverse applications. Their behavior and properties are extensively analyzed to ensure their suitability for tasks requiring high levels of unpredictability and apparent randomness. Building upon these principles, the subsequent section details the development and explicit mathematical formulation of our novel four-stage synchronized chaotic map.

## 2. MAP DEVELOPMENT

The new proposal features four stages working in conjunction. The traditional Bernoulli map [24] operates exclusively in the first quadrant (positive input, positive output), as shown in Figure 1. The proposed map, however, works in all four quadrants. The first stage receives and delivers positive values, similar to the original map.

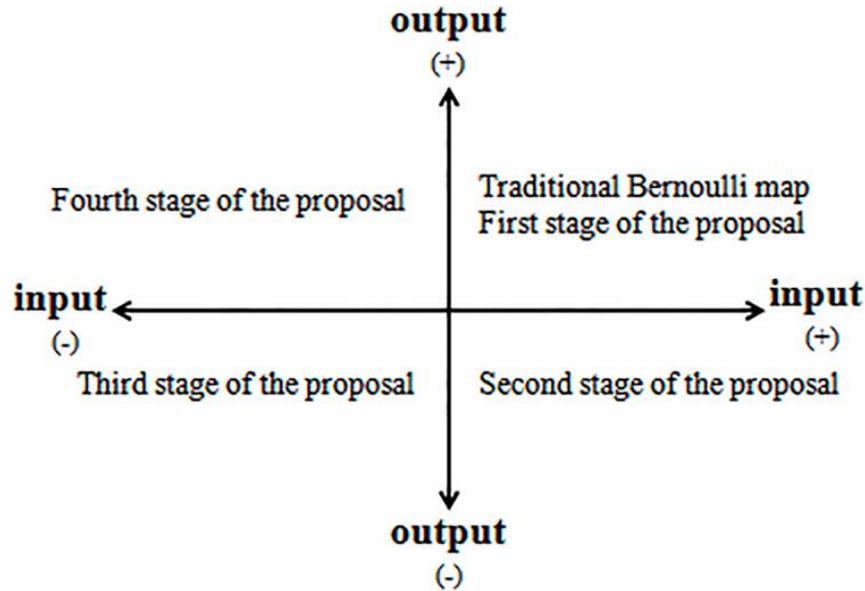

Figure 1. Description of the coordinated system generated by input and output signs

The second stage receives positive values and delivers negative ones, operating in the fourth quadrant. The third stage receives and delivers negative values, working in the third quadrant. Finally, the fourth stage receives negative values and delivers positive ones, operating in the second quadrant and completing the cycle. The proposal operates in an anti-clockwise direction, as depicted in Figure 1.

Equation 1 explicitly defines the proposed synchronized map, describing the evolution of the system through its four operational stages. This equation is fundamental for understanding how the current output state of the map ($x_n$) determines its future state ($x_{n+1}$), and how the feedback parameters ($\mu$) modulate this behavior. The equation is expressed as a system of piecewise functions, where $x_{n+1}$ takes a different form depending on the stage the system is currently in:

$$x_{n+1} = \begin{cases} 1 - (1-\mu_1)[2x_n \, modulo(1)] & first\,stage \\ -(1-\mu_2)[(1-2x_n)\,modulo(1)] & second\,stage \\ \mu_3[2x_n\,modulo(-1)] & third\,stage \\ -\mu_4[(1-2x_n)\,modulo(-1)] & fourth\,stage \end{cases}$$

Equation 1

Where:

- $x_n$ represents the current output state of the map at a given time. This value serves as the input to the current stage of the cycle.
- $x_{n+1}$ denotes the future state of the map, which will become $x_n$ for the next step in the cycle.
- $\mu_{1,2,3,4}$ are the feedback parameters corresponding to the First, Second, Third, and Fourth Stages, respectively. These values are crucial as they control the "strength" of the feedback and, consequently, the dynamic behavior of the map in each quadrant. By adjusting these parameters, one can significantly influence how the system transitions and behaves in each phase.
- The operation X(modY) (modulo) returns the remainder of the division of X by Y. This operation is key for confining values within a specific range and for transitions between quadrants.

Each of the four branches of the equation corresponds to a specific operational stage of the cycle, functioning in a distinct quadrant of the Cartesian plane, as described in the text:

- First Stage: Operates in the first quadrant. It receives and delivers positive values. The expression $1-(1-\mu_1)[2x_n(\bmod 1)]$ models how the positive input $x_n$ is transformed into a positive output, with $\mu_1$ controlling the slope or scaling factor.
- Second Stage: Operates in the fourth quadrant. It receives positive values and delivers negative ones. The formula $-(1-\mu_2)[(1-2x_n)(\bmod 1)]$ ensures that a positive $x_n$ input results in a negative output, with $\mu_2$ adjusting the intensity.
- Third Stage: Operates in the third quadrant. It receives and delivers negative values. The expression $\mu_3[2x_n(\bmod -1)]$ transforms a negative $x_n$ input into a negative output. The modulo with $-1$ is important here to maintain the correct direction for negative values.
- Fourth Stage: Operates in the second quadrant. It receives negative values and delivers positive ones, completing the anti-clockwise cycle. The expression $-\mu_4[(1-2x_n)(\bmod -1)]$ converts a negative $x_n$ input into a positive output.

In summary, Equation 1 is the core of the proposed model. It describes how the system advances from one state to another deterministically, with each stage designed to operate within a specific quadrant of the plane, and the μ parameters offering the necessary control to tune the chaotic or desired dynamics of the map.

Figure 2 illustrates each stage operating in different quadrants, generating a map that functions within the interval (-1, 1). The stages presented in the figure show their characteristic behavior using $\mu_{1,2}=0.01$ for the first and second stages, and $\mu_{3,4}=0.99$ for the third and fourth stages. For stage selection, the map employs a simple mechanism. In the first iteration, the map selects the first stage. For the second iteration, the second stage is selected; then, for the next two iterations, the third and fourth stages are selected, respectively. When the fifth iteration occurs, the process restarts with the first stage, and the cycle repeats indefinitely.

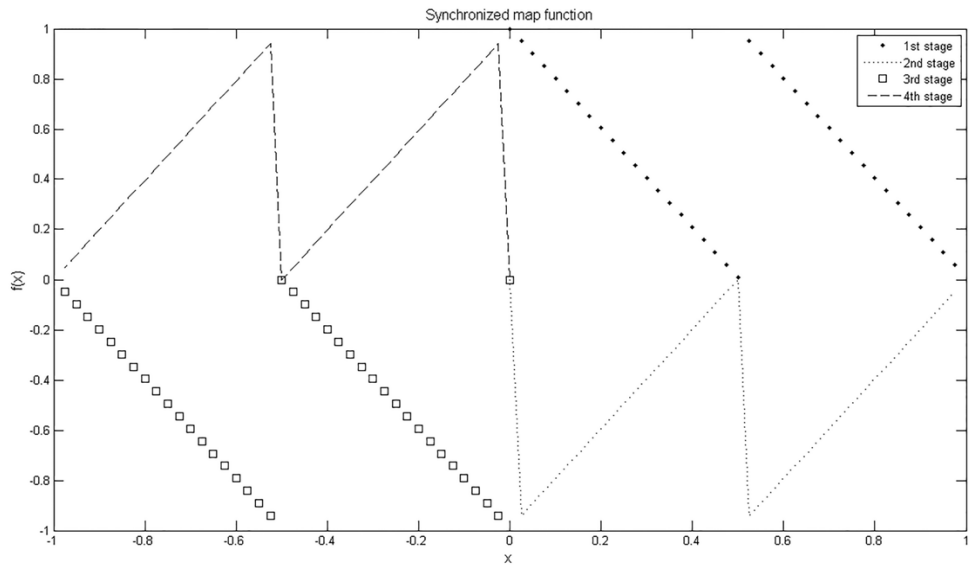

Figure 2. Graphical description of the map described in Equation 1

Using each stage of Eq. (1), different apparently random sequences are generated; one sequence from each stage is presented in the lower part of Figure 3. The most visible feature of each sequence in Figure 3 is its range: the sequence generated by the first stage has a range of (0, 1), the second one is (-1, 0), the third one is (-1, 0), and the fourth stage sequence is (0, 1). By utilizing every stage, the synchronized map generates a sequence with a range of (-1, 1), which is shown in the upper part of Figure 3.

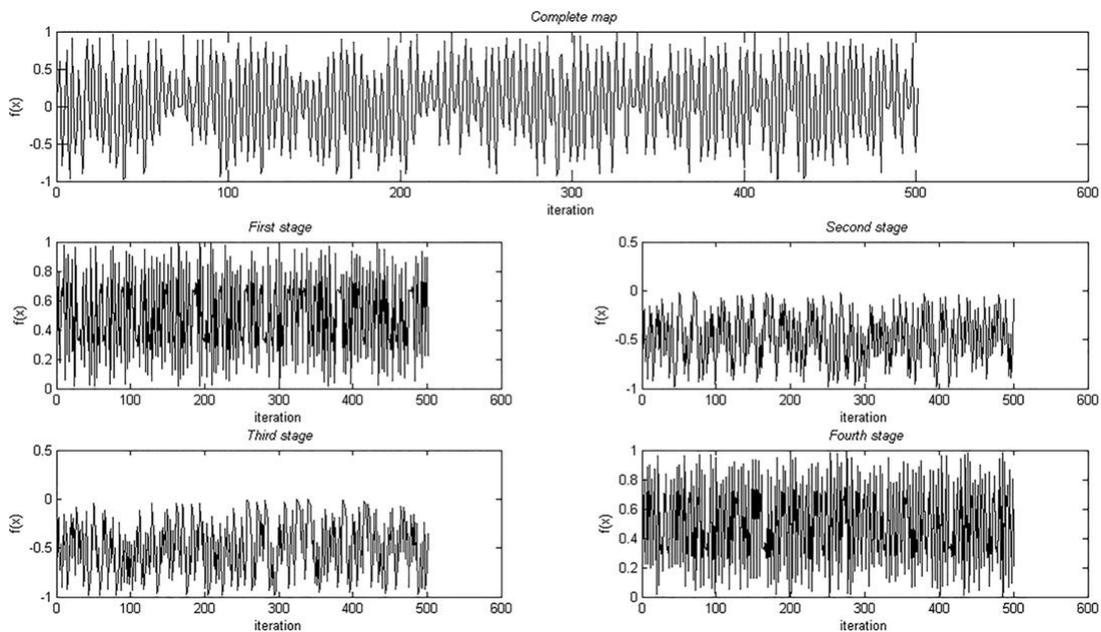

Figure 3. Different sequences obtained from the stages of the map, and the one obtained from the synchronized map

To gain a deeper understanding of the proposed map's dynamic characteristics, we proceed with a detailed graphical analysis using established visualization tools.

## 3. GRAPHICAL ANALYSIS

The cobweb plot is a visual tool that illustrates the qualitative behavior of one-dimensional iterated functions [26]. Devaney referred to this tool as Graphical Analysis [27]; it helps observe the long-term orbit of a specific map under given initial conditions. To construct a cobweb plot, a diagonal line is first drawn. The map is then iterated as many times as required. Once the orbit has been determined, a vertical line is traced from coordinate $(0, x_0)$ to $(x_1, x_0)$; subsequently, a horizontal line is traced from $(x_1, x_0)$ until it reaches $(x_1, x_1)$. After this initial trace, the process is repeated for each iteration, following: vertical traces from $(x_{n-1}, x_{n-1})$ to $(x_n, x_{n-1})$ and horizontal traces from $(x_n, x_{n-1})$ to $(x_n, x_n)$.

Based on the previously described procedure, the cobweb for each stage was plotted; the results appear in Figure 4. The plots for the first and fourth stages are quite similar, both having two sinking points. In contrast, the second and third stage plots do not show any attractors; they appear to be similar projections.

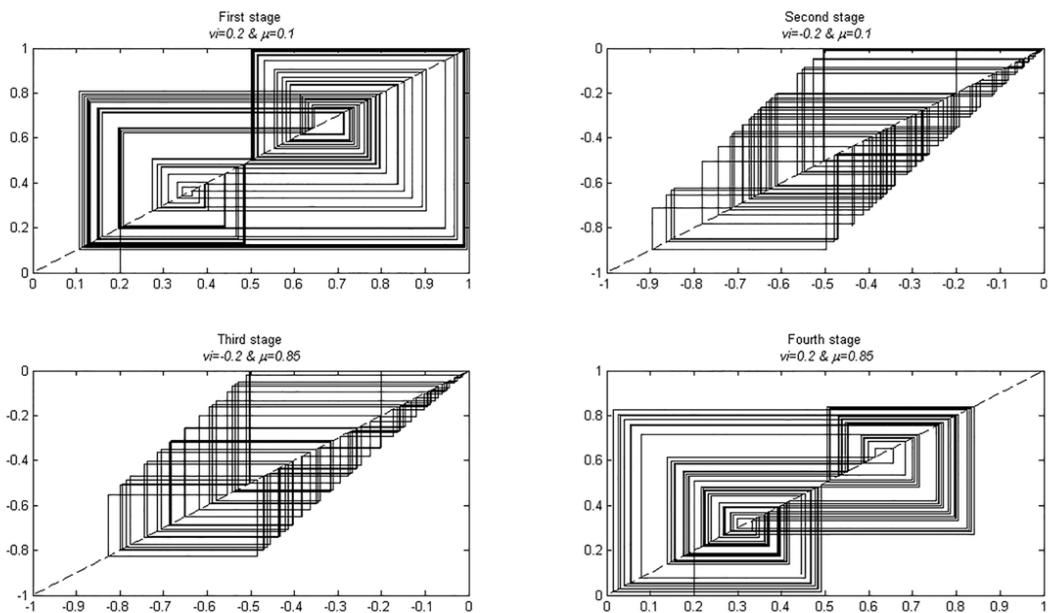

Figure 4. Each individual stage cobweb plot

The basic cobweb determination procedure is suitable for simple functions. However, in the present work, we propose a four-synchronized stages map; therefore, we need to propose a modified cobweb determination procedure. This procedure is quite similar to the basic one; it simply requires expanding the x-axis range from (0, 1) to (-1, 1) and the y-axis also from (0, 1) to (-1, 1), thereby increasing the number of quadrants.

The modified procedure works to represent the new map's orbits; in Figure 5, the cobweb plot for the proposed map is presented. The first quadrant cobweb is generated by the first stage function, the second quadrant by the fourth stage function, the third quadrant by the third stage function, and the fourth quadrant by the second stage function. The new map inherits certain behaviors from its progenitor individual stage functions; the first and third quadrant cobwebs are quite similar to their respective individual stage cobwebs. On the other hand, the second and fourth quadrant cobwebs are affected.

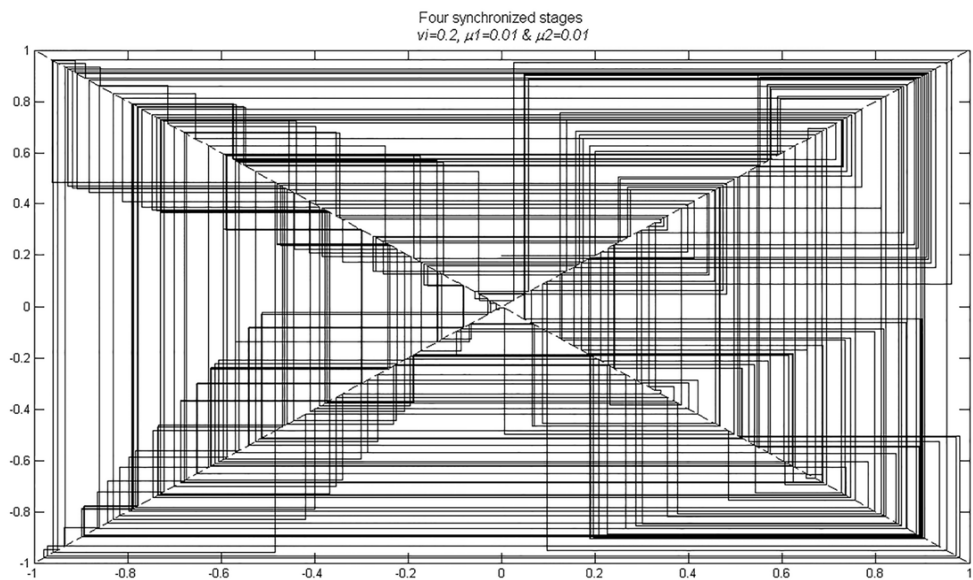

Figure 5. Cobweb plot of the newly proposed four synchronized stages map

Beyond qualitative visualizations like cobweb plots, bifurcation diagrams offer a crucial quantitative perspective on a system's long-term behavior as its parameters vary. The following section presents a comprehensive analysis using these diagrams.

## 4. BIFURCATION DIAGRAMS

These diagrams illustrate the long-term values of a mathematical function given specific initial values [28]. The term 'bifurcation' refers to a system's ability to produce several different output values; traditionally, the parameter that controls the range of output values is the feedback parameter, also known as µ [29]. Figure 6 shows four bifurcation diagrams, one for each stage. The first and fourth stage diagrams are similar but mirrored, as are the second and third stage diagrams.

Due to their similarity, the first and second stages will be used as references. The first stage diagram is dense for µ∈ (0,0.2). After that, the diagram is divided into two zones for µ∈ (0.2,0.35) and then into four zones for µ∈ (0.35,0.5). Finally, the bifurcation diagram has only one possible output, indicating that the stage is no longer chaotic. The fourth stage diagram exhibits the same behavior, but the ranges are mirrored.

The second stage diagram is dense for µ∈ (0,0.5); however, the possible output range diminishes from -1 to -0.5. When µ>0.5, the diagram shows a complete absence of output for the function; in fact, the system output for µ∈ (0.5,1) is almost 0. The third stage diagram has similar behavior but starts with values closest to 0 and finishes dense and covering the complete available output range (-1, 0).

International Journal of Chaos, Control, Modelling and Simulation (IJCCMS) Vol.14, No.1/2, June 2025

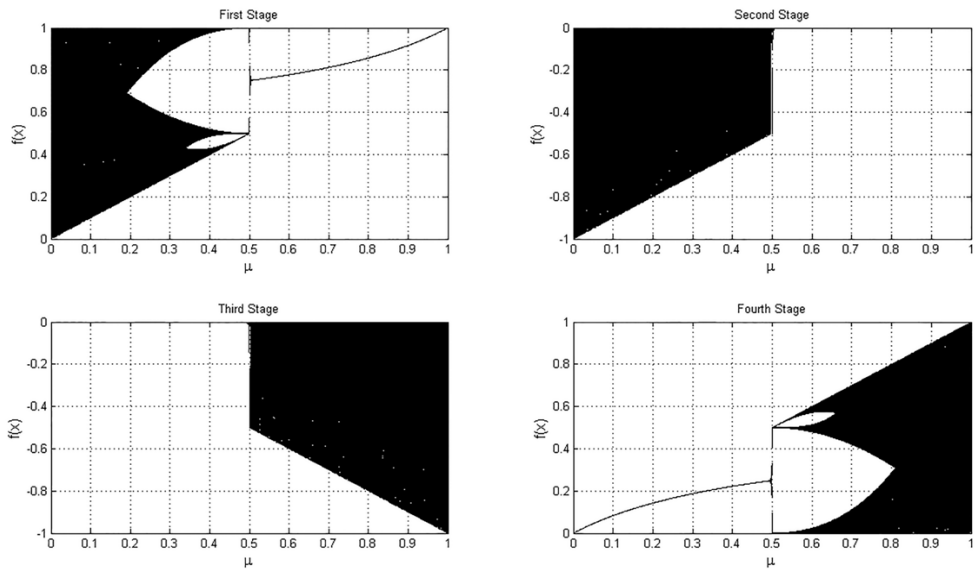

Figure 6. Bifurcation diagrams for each independent stage

As previously mentioned, bifurcation depends on the feedback factor; therefore, this factor must be varied constantly and uniformly, which is not possible since the proposed map has four functions and requires some arrangements. The first arrangement is to unify the values; by unifying them, the bifurcation diagram in Figure 7 is elaborated.

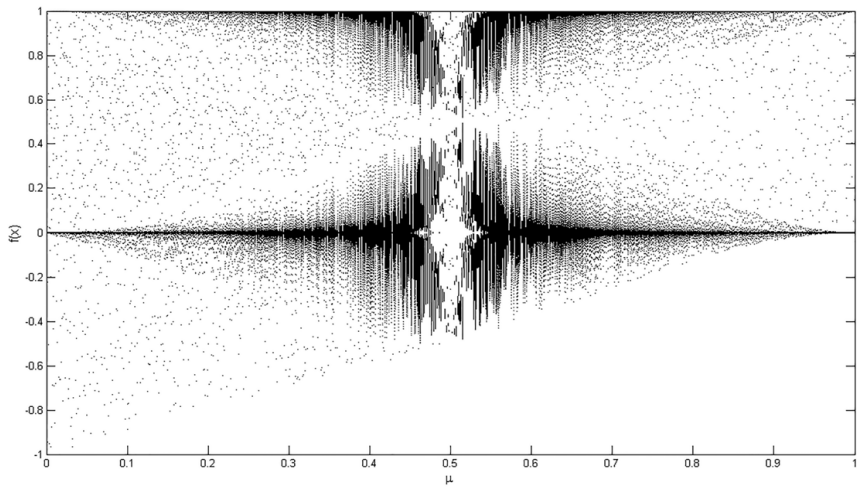

Figure 7. Failure to obtain the bifurcation diagram of the proposed map

The diagram in Figure 7 represents a failure because using the same μ value for all four functions compromises some outputs, causing the map to not work properly. Figure 7 demonstrates that the same μ value cannot be used for every stage.

To generate a correct diagram, another modification is required. The second modification is to propose individual feedback factors: $\mu_1$, $\mu_2$, $\mu_3$, and $\mu_4$. These individual factors correspond to each individual stage; however, the bifurcation diagram is defined for only one feedback factor. Therefore, another modification is required; this time, let us define $\mu=\mu_1=\mu_2$ and $\mu_3=\mu_4=(1-\mu)$. Using this third modification, the bifurcation diagram in Figure 8 was obtained.

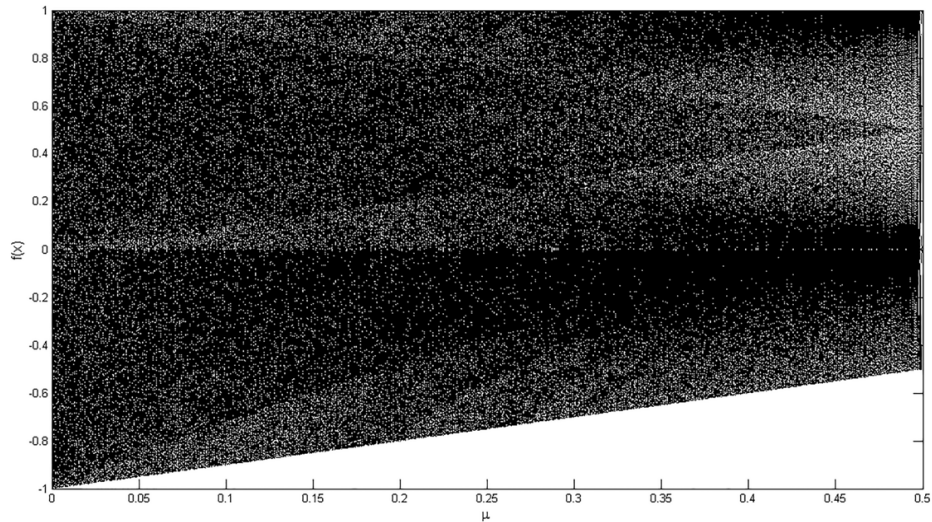

Figure 8. Bifurcation diagram for the proposed map

Since the third modification, the μ range of the obtained bifurcation diagram is (0, 0.5); nonetheless, the system output range is (-1, 1), covering the entire available space. The diagram has a silhouette similar to the second stage diagram, disregarding its shadow zones. They appear as triangle-shapes; two of them are easily recognized, starting at μ=0 and ending at μ=0.5. Additionally, there are other shadow zones, but they are blurrier than the triangles, making them difficult to identify.

Having characterized the qualitative and quantitative dynamics of the proposed map, we now present the empirical results of its statistical validation, focusing on the rigorous NIST testing suite.

## 5. RESULTS

For presentation, the results focus on the NIST statistical testing suite with a sequence generated using the complete map.

### 5.1. NIST Tests

A Matlab script was developed using the mathematical descriptions for the proposed map. This script produced a 100MB TXT file containing sequences, which was used to test the map's behavior. The control parameters used for file creation were selected to offer high concordance across all our presented results. Table 1 shows two columns: one for the proposed Bernoulli map and another for the classic Bernoulli map. The classic Bernoulli map has the following mathematical description:

$$X_{n+1} = \mu[2x_n \bmod(1)], \qquad \text{Equation 2}$$

where $\mu \in (0,1)$ and $x_n \in (0,1)$. Equation 2 was transformed to $\mu \in (0, 2a)$ and $x_n \in (0, 2b)$, yielding the next expression:

$$X_{n+1} = \frac{\mu}{2^b} \left[ 2x_n \bmod (2^a) \right]$$

Equation 3

With the mathematical definition for the classic Bernoulli map within the proper limits, a 100 MB TXT file with sequences was generated. The initial value for the generation was 858,993,459, and the feedback value was 224; these values are equivalent to those used for file generation with the proposed map. Everything was defined to establish the fairest possible comparison between both maps.

According to Table 1, the proposed map outperforms the classic one in the Frequency, Cumulative Sums, Runs, and FFT tests. It shows the same result for Longest-run but is slightly inferior in the Rank test. Despite the comparison, the proposed map passed all its tests, while the classic one did not.

Table 1: Comparison table for Classic Bernoulli and Proposed maps

| Statistical test | Classic Bernoulli Map | Proposed Map |
| --- | --- | --- |
| Frequency | 0.5781 | 0.9893 |
| Cumulative-sum 1 | 0.5458 | 0.9905 |
| Cumulative-sum 2 | 0.5690 | 0.9892 |
| Runs | 0.8257 | 0.9444 |
| Longest-run | 1.0000 | 1.0000 |
| Rank | 0.9935 | 0.9907 |
| FFT | 0.9845 | 0.9992 |

## 6. CONCLUSIONS

A new chaotic map was proposed. Despite its individual stages, they operate as a single unified system. To join the individual maps, several modifications were performed. While the traditional Bernoulli map is defined from 0 to 1, the new one is defined from -1 to 1. The proposed map's behavior was validated by calculating its bifurcation diagram and cobweb plot. To calculate the bifurcation diagram, some adjustments were needed; the four feedback factors in the proposed map were related. This adjustment was solely for diagram calculation; in a real application, they can be independent. After the bifurcation diagram, the cobweb plot was calculated. Traditional cobweb plotting relates a map's current iteration result to its immediate next iteration; however, additional features were needed to represent the real behavior of our proposed map.

After minor modifications, the Matlab script produced a 100MB TXT file, which was used to demonstrate the statistical quality of the random sequences generated by the proposed map. The tests were performed with a suite released by the NIST. To provide a reference frame, another 100MB TXT file was created using the traditional Bernoulli map and tested with the same suite. In contrast to the original map, the proposed one passed all tests. These promising statistical outcomes lead us to the following conclusions regarding the design and potential applications of our novel chaotic map.

The inherently robust design of the proposed map, particularly its synchronized multi-stage piecewise linear architecture, holds significant implications for its digital implementation. Unlike many chaotic systems that experience substantial degradation when realized with finite precision, our map demonstrates resilience, successfully maintaining its chaotic properties and passing rigorous statistical tests. This sustained performance in a digital environment is crucial for practical applications where computational constraints or the need for high-quality randomness are paramount. Its design suggests a promising avenue for developing reliable hardware-based chaotic generators for cryptographic purposes, secure communications, and other embedded systems where the integrity and unpredictability of chaotic sequences are critical for security and system stability.

## AUTHOR

**Ricardo Francisco Martínez-González** , He was born in Veracruz, Veracruz, Mexico, on January 2, 1983. He earned his Bachelor's degree in Electronic Engineering from the Instituto Tecnológico de Veracruz. He pursued his Master's and Ph.D. degrees in Science, specializing in Electronics, at the Instituto Nacional de Astrofísica, Óptica y Electrónica (INAOE). He has worked as a software platform and mobile application developer for private companies. Currently, he works at the Tecnológico Nacional de México campus Veracruz, in the Department of Electrical-Electronic Engineering, where he has directed various research projects in areas such as artificial vision, control, instrumentation, artificial intelligence, steganography, energy efficiency, among others. He is the author of several scientific articles in the aforementioned areas. 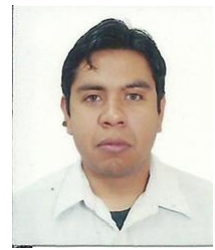